\documentstyle[preprint,prb,aps]{revtex}

\begin{document}
\title{{\bf Statics and dynamics of charge fluctuations in the t-J model}}  
\vspace{2cm}
\author{{\bf Roland Zeyher}$^a$ and {\bf Miodrag L. Kuli\'c}$^{a,b}$}

\address{$^a$Max-Planck-Institut\  f\"ur\
Festk\"orperforschung,\\ Heisenbergstr.1, 70569 Stuttgart, Germany \\
$^b$Universit\"at Stuttgart, 2. Physikalisches Institut,\\ 70550
Stuttgart,Germany}

\date{\today }

\vspace{3cm}

\maketitle

\begin{abstract}
The equation for the charge vertex $\gamma$ of the $t-J$ model is
derived and solved in leading order of an 1/N expansion, working
directly in terms of Hubbard operators. Various
quantities which depend crucially on $\gamma$ are then calculated,
such as the life time and the transport life time of electrons
due to a charge coupling to other degrees of freedom and the
charge-charge correlation function. Our
results show that the static screening of charges
and the dynamics of charge fluctuations depend only weakly on $J$
and are mainly determined by the constraint of having no double
occupancies of sites.
\end{abstract}

\newpage 

The charge fluctuation spectrum and the screening properties
of the t-J model are largely determined by the charge
vertex $\gamma(k,q)$. Here, $k$ denotes both the momentum $\bf k$
of an electron and the Matsubara frequency $i \omega_n$.
Similarly, $q$ stands for the transferred momentum $\bf q$
and the frequency $i \nu_n$. A simple interpretation of $\gamma$
has been given in \cite{Kul1}: It multiplies the bare electron-phonon
interaction $g(k,q)$ to yield an effective electron-phonon
interaction which takes into account screening effects due to
the constraint of no double occupancies of sites. Another property
of $\gamma$ is that its poles in the second frequency argument  
determine the dispersion of collective density waves and thus is
an important ingredient for the density-density correlation
function $D(q)$. For the case $J=0$ the properties of the
static vertex $\gamma$ has been investigated in \cite{Kul1,Zey1}.
In particular, it has been shown that $\gamma$ exhibits a strong momentum
dependence in $\bf q$ at low frequencies and small and intermediate
dopings which tends to suppress the effective electron-phonon
interaction in the $tt'$-model ($t$ amd $t'$ denote hopping
integrals between nearest and second-nearest neighbor sites,
respectively). This effect is especially pronounced in transport
quantities. The frequency dependence of $\gamma$
and $D$ has been investigated, again for $J=0$, in \cite{Geh1}.
There it has been shown that $D$ is nearly exclusively determined
by collective effects and has an energy scale substantially larger than 
the effective band width, in agreement with computer simulations \cite
{Toh1}. The purpose of this communication is to extend the above results
to the t-J model and to investigate to what extent the above
properties of $\gamma$ and $D$ depend on $J$. 

The Hamiltonian of the t-J model reads
\begin{equation}
H = -\sum_{{ij} \atop {p=1...N}} {{t_{ij}} \over N} X_i^{p0} X_j^{0p}
+ \sum_{{ij} \atop {p,q=1...N}} {{J_{ij}} \over {4N}} X_i^{pq}
X_j^{qp} .
\end{equation}
The subscripts $i,j$ stand for lattice sites; the superscripts
$p,q$ denote for $p=0$ the unoccupied and for $p=1...N$ singly
occupied states with a spin index $p$. This means that 
the original $SU(2)$ spin space has been extended to a $SU(N)$ space 
which is a
well-known procedure in slave boson calculations \cite{Gri1}.
The Hubbard operators $X_i^{pq}$ with $p=0,q=0$
and $p>0,q>0$ have bosonic and those with
$p=0,q>0$ or $p>0,q=0$ have fermionic character. They obey the
following commutation and anticommutaion rules, respectively,
\begin{equation}
[X_i^{pq}, X_j^{rs}]_{\mp} = \delta_{ij}(\delta_{qr} X_i^{ps} \mp
\delta_{sp} X_i^{rq}).
\end{equation}
In the $SU(N)$ model, considered here, the $X$-operators are
subject to the constraint
\begin{equation}
\sum_{p=0}^N X_i^{pp} = {N \over 2}.
\end{equation}
This means that at most $N/2$ of the $N$ states at each site can be
occupied at the same time. The first term in Eq.(1) describes
the hopping of particles between the sites $i$ and $j$ with 
matrix elements $t_{ij}$. The second term in Eq.(1) denotes
the Heisenberg interaction between the spin densities at site $i$ and
$j$ with the exchange constants $J_{ij}$. In the following we 
consider $J_{ij}$ only between nearest neighbors ($J_{ij}=J$)
and $t_{ij}$ between nearest ($t_{ij}=t$) and next nearest
($t_{ij}=t'$) neighbors. The coupling constants in Eq.(1)
have been scaled with $N$ in such a way that the limit
$N\rightarrow \infty$ describes an interesting physical case and
that for $N=2$ the usual t-J model is, except for an overall factor $1/2$,
 recovered. \par
Using a $1/N$ expansion the Hamiltonian Eq.(1) has been
investigted for $J=0$ in \cite{Kul1} and, in more detail, in \cite{Zey1}.
These treatments can be generalized to a finite value of $J$
in a straightforward way: The equation of motion for a fermionic
X-operator is, using Eq.(1):
\begin{equation}
{{\partial} \over {\partial \tau_1}} X^{0q}(1) = \sum_{p_2,q_2,q_3}
\int d{2} d{3} t({{0p_1} \atop {1}} {{p_2q_2} \atop
{2}} {{0q_3} \atop {3}}) X^{p_2q_2}(2) X^{0q_3}(3)
\end{equation}
with 
\begin{eqnarray}
t({{0q_1} \atop {1}} {{p_2q_2} \atop{2}} {{0q_3} \atop
{3}})& =& \delta({1}-{2}) \delta_{p_20} \delta_{q_20}
\delta_{q_1q_3}t({1}-{3})/N  \nonumber \\
+&&\delta_{p_2q_3} \delta_{q_1q_2}
(\delta({1}-{2})t({1}-{3}) - \delta({1}
-{3})J({1}-{2})/2)/N.
\end{eqnarray}
Here, ${1}$ is an abbreviation for $i_1 \tau_1$, i.e.,
${1}=(i_1 \tau_1)$, where $\tau_1$
denotes the imaginary time. $t({1}-{2})$ is equal to
$t_{i_1i_2}\delta(\tau_1-\tau_2)$ and $J({1}-{2})$
equal to $J_{i_1i_2} \delta(\tau_1-\tau_2)$. The first term in the
parantheses on the right-hand side of Eq.(5) describes hopping without
flip of the spin whereas the second one hopping with a spin-flip.
Comparing the above Eqs.(4) and (5) with Eqs.(9) and (10) of 
Ref.\cite{Zey1}
one finds that the Heisenberg term in $H$ just adds a contribution to the
spin-flip hopping term. The perturbation expansion in \cite{Zey1}
rests on two relations: The equation of motion for fermionic
Hubbard operators and Eq.(31) in \cite{Zey1} which relates expectation
values of bosonlike Hubbard operators to Green's functions. The
first relation is modified by the Heisenberg term in the above way,
the second
relation is unchanged. As a result, it is straightforward to generalize
the expressions for the self-energy, the vertex etc. in \cite{Zey1} to the
case of a finite $J$. \par
Using the above procedure one obtains from Eq.(37) in \cite{Zey1}
the following expression for the self-energy in $O(1)$
of the t-J model:
\begin{equation}
\Sigma({1}-{2}) = \delta({1}-{2})
\int d{3} t({1}-{3}) g({3}^+ -{1}) 
-t({1}-{2}) <X^{00}({1})> -{ {J({1}^+-{2})}\over 2}
g({1}-{2}).
\end{equation}
The normalized Green's function $g$ (denoted by $\tilde G$ in
\cite{Zey1}) is related to $\Sigma$ via Dyson's equation
\begin{equation}
\int d{3} \Big( -\delta({1}-{3}) {\partial \over
{\partial \tau_1}} - \Sigma({1}-{3}) \Bigr) 
g({3}-{2}) = \delta({1}-{2}).
\end{equation}
$<X^{00}({1})>$ is the expectation value of $X^{00}({1})$.
Both, $\Sigma$ and $g$ are diagonal in the internal indices so we have
omitted them in the above equations. The self-energy in Eq.(6)
is instantaneous giving rise to a frequency-independent, but
momentum-dependent renormalized one-particle energy $\epsilon({\bf k})$.
After a Fourier transformation Eqs.(6) and (7) yield $g({\bf k},i
\omega_n) = 1/(i \omega_n - \xi({\bf k}))$ with
\begin{equation}
\epsilon({\bf k}) = \Delta -q_0 t({\bf k}) -{1 \over {2N_c}}
\sum_{\bf p} J({\bf k}+{\bf p}) f(\xi({\bf p})).
\end{equation}
Here we have $\xi({\bf k}) = \epsilon({\bf k}) - \mu$ and
$\Delta = {1 \over N_c} \sum_{\bf p} t({\bf p}) f(\xi({\bf p}))$,
where $f$ is the Fermi function, $N_c$ the number of primitive cells
and $q_0=\delta/2$ with the doping $\delta$. \par
Taking the Heisenberg interaction also into account the
vertex equation (39) in \cite{Zey1} becomes
\begin{eqnarray}
{\tilde \Gamma}({1}{2};{3})   =
 \delta({1}-{2})
\delta({1}-{3}) + t({1}-{2}) \int d{5} 
d{6} g({1}-{5}) {\tilde  \Gamma}({5}{6};
{3}) g({6}-{1}^+) \nonumber \\
+\delta({1}-{2}) \int d{4}d{5}d{6}
t({1}-{4})g({4}-{5}) {\tilde \Gamma}(
{5}{6};{3}) g({6}-{1}) -{{J({1}-{2})}
\over 2}
\int d{5}d{6} g({1}-{5}) {\tilde \Gamma}({5}
{6};{3}) g({6}-{2}).
\end{eqnarray}
Writing ${\tilde \Gamma}({1}{2};{3})=\gamma({1}
-{2},{1}-{3})$ Eq.(9) becomes after a Fourier 
transformation 

\begin{equation}
\gamma(k,q) = 1 +{{T}\over N_c} \sum_{k'} (t(k)+t(k'+q)-{{J(k-k')}\over 2})
g(k')g(k'+q) \gamma (k',q).
\end{equation}
Since the $\bf k$-dependence of $J$ and $t$ are given by 
trigonometric or products of trigonometric functions Eq.(10)
represents an integral equation with a kernel consisting of 6
separable contributions. Eq.(10) thus can be reduced to a 6x6 
matrix equation with the solution
\begin{equation}
\gamma(k,q) = 1 -\sum_{\alpha=1}^6 F_\alpha({\bf k}) \sum_{\beta=1}^6
(\delta_{\alpha \beta} + \chi(q))^{-1}_{\alpha \beta} \chi_{\beta 2}
(q),
\end{equation}
\begin{equation}
\chi_{\alpha \beta}(q) = \sum_{k'} G_\alpha (k',q) F_\beta (k').
\end{equation}
The vectors $F$ and $G$ are given by
\begin{equation}
F_\alpha({\bf k}) = (t({\bf k}),1,Jcosk_x,Jsink_x,Jcosk_y,Jsink_y)
\end{equation}
and
\begin{equation}
G_\alpha(k,q)=(1,t({\bf k}+{\bf q}),cosk_x,sink_x,cosk_y,sink_y)
\Pi(k,q)
\end{equation}
with $\Pi(k,q)=-g(k)g(k+q)$. The frequency sum in Eq.(12) involves
only $\Pi$ and can easily be carried out:
\begin{equation}
\sum_{n'} \Pi(k',q) = {{f(\xi({\bf q}+{\bf k'}))-f(\xi({\bf k'})} \over
{\xi({\bf k'})-\xi({\bf q}+{\bf k'}) -i \nu_n}}.
\end{equation}
Calculation of $\gamma(k,q)$ thus requires essentially the
calculation of the susceptibility matrix $\chi_{\alpha \beta}(q)$.
For $J=0$ the matrix inversion in Eq.(11) can be done explicitly
and one obtains Eqs.(12)-(15) of \cite{Zey1}. In contrast to this
special case, $\gamma(k,q)$ depends for $J \neq 0$ also 
on the vector ${\bf k}$ for a given doping. \par
The Green's function $\tilde{D}(q)$ for density fluctuations is given by
\begin{equation}
\tilde{D}(q) = {T \over V} \sum_k \gamma(k,q) \Pi(k,q).
\end{equation}
Using Eq.(11) we obtain
\begin{equation}
\tilde{D}(q) = -\sum_{\beta=1}^6 (1+\chi(q))^{-1}_{1 \beta} \chi_{\beta 2}(q).
\end{equation}
Carrying out the analytic continuation $i\omega_n \rightarrow \omega
+i \eta$ the density-density correlation function $D({\bf q},\omega)$
is equal to the negative imaginary part  
of $\tilde{D}({\bf q},\omega+i \eta)$.

We have evaluated numerically the susceptibility matrix 
$\chi_{\alpha \beta}(q)$ using a typical mesh of 1000x1000 points
in the Brillouin zone for the summation over $\bf k'$. Fig.1 shows
results for the zero-frequency vertex $\gamma({\bf k},{\bf q})$
for $J=0.2, t'=0$ (all energies are in units of the nearest-neighbor
hopping energy t) and three different dopings $\delta$.
For each doping ${\bf k}$ is put on the Fermi line in $(1,1)$-direction
and $\bf q$ is varied along the $(1,1)$-direction from zero
to the maximal momentum transfer for points on the Fermi line. Fig.1 should
be compared with Fig.1 of$^1$ where a similar plot for $\gamma$
is given for the case $J=t'=0$. (The momentum scale in that reference 
should be scaled by a factor $\sqrt{2}$ in order to have the same
absolute scale). The Figures clearly show that the
main property of the vertex found in $^1$ for $J=0$ is also present 
for $J=0.2$: For large dopings $\gamma$ varies only smoothly with 
momentum whereas at smaller dopings $\gamma$ develops a forward scattering
peak with a width $\sim \delta$ due to a strong suppression of
large momentum transfers. This implies that the effective charge interaction
of electrons with other degrees of freedom (impurities, phonons etc.)
is essentially the bare one at small
but heavily suppressed at large momenta.
The curve for $\delta=0.10$ in Fig.1 shows a new feature: It passes
through zero at a small momentum which means that the effective
interaction is exactly zero at this point due to correlation effects. 
With increasing
momentum it goes through a minimum with a negative value and approaches
zero from below at large momenta. A similar, but less pronounced
behavior, has been found \cite{Zey1} in the one-dimensional $t$-model 
and in the
two-dimensional $tt'$-model with a finite $t'$. Finally, we have
chosen in Fig.1 a rather small value for $J$ and not too small values
for $\delta$ in order to avoid singularities in $\gamma$ due to
instabilities of the homogenous phase$^5$. \par
The above vertex function allows to answer the following
question: How much are the inverse life time $1/\tau$ and inverse
transport life time $1/\tau_{tr}$ of an electron affected by
electronic correlations if the coupling of the electron to additional
degrees of freedom (phonons, impurities etc.) is due to the interaction
of charges? The answer becomes especially simple if one assumes
that the bare coupling function is structureless, i.e., is independent
of momentum and frequency. In \cite{Zey1} is has been shown that the 
quantities
$\Lambda_1$,$\Lambda_{tr}$ defined by
\begin{equation}
\Lambda_1 = C << { {|\gamma({\bf k},{\bf k}-{\bf k'})|^2}\over q_0}
>_{\bf k}>_{\bf k'},
\end{equation}
\begin{equation}
\Lambda_{tr} = C<< { {|\gamma({\bf k},{\bf k}-{\bf k'})|^2}\over q_0} (
{\bf v}({\bf k})-{\bf v}({\bf k'}))^2>_{\bf k}>_{\bf k'}/
(2<<{\bf v}^2({\bf k})>_{\bf k}>_{\bf k'},
\end{equation}
describe changes in the inverse life time $1/\tau$ (or, in the Eliashberg 
function
$\alpha^2F(\omega)$ for s-wave supercondcutivity) and the inverse
transport life time $1/\tau_{tr}$ due to correlation effects.
The overall factor $C$ is chosen such  that $\Lambda_1=\Lambda_{tr}=1$
for $\delta \rightarrow 1$, i.e., the empty band limit. If $\gamma$
in Eqs.(18) and (19) depends only weakly on momentum we have
$\Lambda_1 \sim \Lambda_{tr}$. On the other hand, if $\gamma$
is nonzero only for ${\bf k} \sim {\bf k'}$ $\Lambda_{tr}$ is much
smaller than $\Lambda_1$. Fig. 2 shows $\Lambda_1$ and $\Lambda_{tr}$
as a function of $\delta$ for $J=0.3$ and $t'=-0.25$. With decreasing doping
$\Lambda_1$ and $\Lambda_{tr}$ first pass through a maximum at around
$\delta \sim 0.8$ and then decrease monotonically by around a factor
2 and 4, respectivley, until $\delta \sim 0.2$. The more and more
pronounced appearance of a forward scattering peak in $\gamma$
at still smaller dopings would cause a further decrease in $\Lambda_1$
and $\Lambda_{tr}$ and especially in the ratio $\Lambda_{tr}/\Lambda_1$.
However, we exclude this low-doping region from our considerations 
because of the occurrence of instabilities of the
homogenous phase in that region. Figure 2 suggests
that correlation effects suppress
$\Lambda_1$ and, even stronger, $\Lambda_{tr}$,
moving from the overdoped 
towards the maximal doped regime. 
Fig. 10 in \cite{Zey1} presents 
results for $\Lambda_1$ and $\Lambda_{tr}$ and for (using our energy 
units) $J=0$ and $t'=-0.20$. Comparing this Figure with our present
Figure 2 one concludes
that $\Lambda_1$ and $\Lambda_{tr}$ depend only very weakly on
$J$. (The additional interpretation of the quantity $\delta \cdot
\Lambda_{tr}$ as being proportional to the resistivity in \cite{Zey1}
should be dropped since the Drude weight entering the static part of
the resisitivity depends strongly on $\delta$ and the resisitivity near
half-filling is characterized by $\Lambda_{tr}/\delta$ rather than by
$\Lambda_{tr} \cdot \delta$). \par
Fig. 3 shows the density-density correlation function $D({\bf q},\omega)$
for $J=0.1$ (left panel) and $J=0.3$ (right panel) for various
momenta ${\bf q}$. Curves corresponding to the same momentum 
practically coincide with each other demonstrating
that $D({\bf q},\omega)$ and thus also the dynamic 
part of the vertex are nearly independent of $J$. This implies,
that $D$ is dominated by collective effects in form of an infinitely sharp,
dispersive
sound peak also in the presence of the Heisenberg interaction.
This peak has been broadened in Fig.3 by using a finite
value of 0.1 for $\eta$.
The energy of this peak
is in general much larger than the width of the renormalized band (
which is 0.96 for $J=0.1$ and 1.28 for $J=0.3$). The contribution of
the particle-hole contimuum to $D$ is nearly invisible if the
sound peak is well above the particle-hole spectrum like, for instance,
in the case ${\bf q}=(\pi,\pi)$. In the other cases like 
${\bf q}=(2\pi/5,\pi/5), (\pi/5,3\pi/5), or (\pi,0)$ the collective
peak is not so well separated from the particle-hole continuum and
$D$ has structure also at low frequency reflecting density of states
of single particle-hole excitations. The absence of an noticeable 
dependence of the peak position on $J$ as well as the quite different
energy scales for charge and spin fluctuations \cite{Geh1} remind
of spin-charge separation found in one-dimensional models. 

In conclusion, the equation for the charge vertex $\gamma$ of the
$t-J$-model has been derived in leading order of an 1/N expansion,
reduced to a 6x6 system of linear equations, and solved numerically.
Our results
for the momentum and frequency dependence of $\gamma$ show only a 
weak dependence
on $J$. We also discussed various properties which depend 
sensitively on $\gamma$,
namely, the effect of correlations on the inverse life time
and the inverse transport life time of electrons and the dynamics
of charge fluctuations. Our conclusion is that these quantities depend
only weakly on $J$ and are mainly
determined by the constraint of having no double occupancies of sites.
These findings
are consistent with recent exact numerical results from small
clusters \cite{Toh1,Pre1}.

{\bf Acknowledgement:} One of us (M.L.K.) would like to thank Prof. 
Michael Mehring for support.

\begin{figure}
\protect
\caption{
Zero-frequency vertex function $\gamma({\bf k},{\bf q})$
as a function of $aq$ with $\bf k$ fixed on the Fermi line along
the $(1,1)$-direction for three dopings $\delta$.}
\end{figure}


\begin{figure}
\protect
\caption{  
Correlation induced enhancements $\Lambda_1$ and $\Lambda_{tr}$
as a function of doping $\delta$ for $J=0.3$ and $t'=-0.25$.}
\end{figure} 


\begin{figure}
\protect
\caption{  
Density-density correlation function $D({\bf q},\omega)$
as a function of energy $\omega$ for $\delta = 0.2$ and $J=0.1$
(left panel) and $J=0.3$ (right panel) using $\eta =0.1$.}
\end{figure}

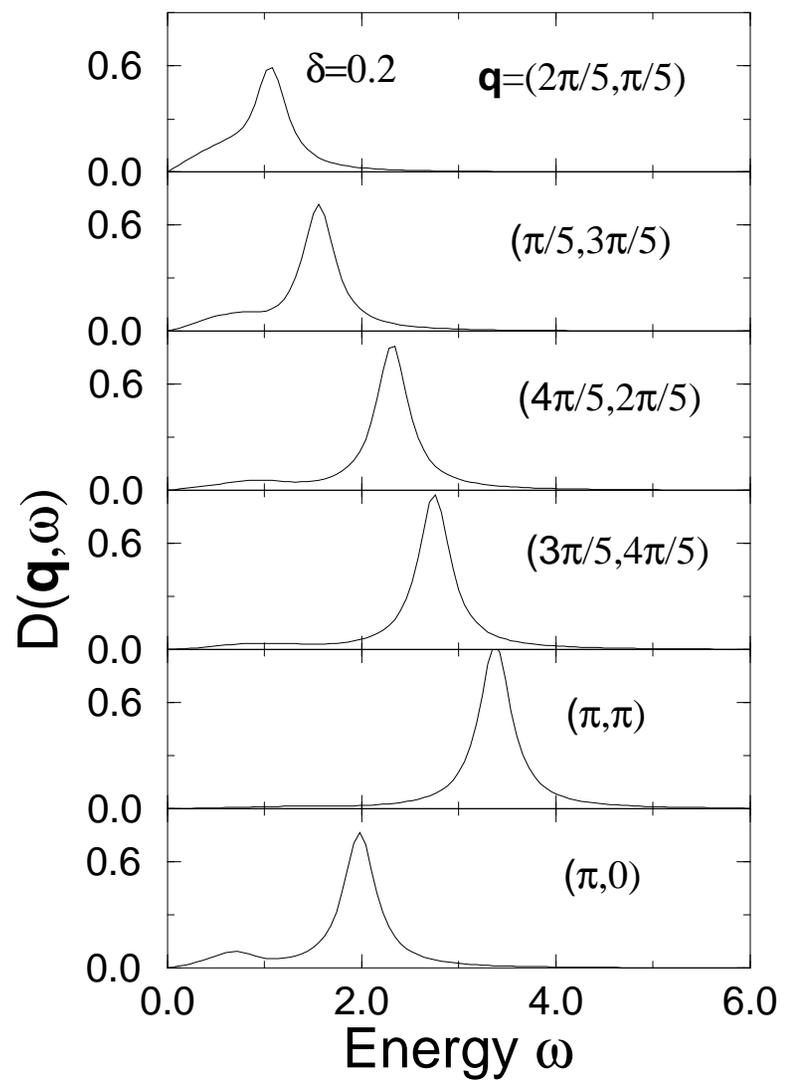
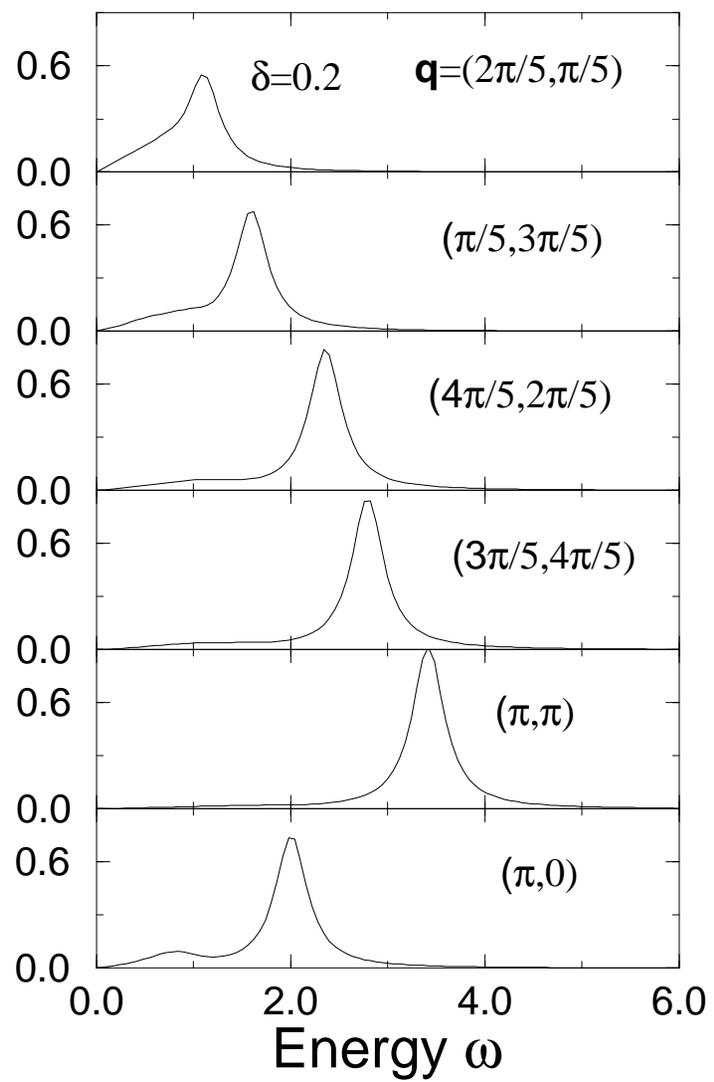
\end{document}